\documentclass[sn-mathphys]{sn-jnl}

\usepackage[T1]{fontenc}
\usepackage{xcolor}
\usepackage{xfrac}
\usepackage{amsmath}
\usepackage{upgreek}
\usepackage{enumerate}

\definecolor{midblue}{rgb}{0.0, 0.4, 0.7}
\definecolor{mypurple}{rgb}{0.7, 0.3, 0.8}

\definecolor{PineGreen}{HTML}{008B72}
\definecolor{Berry}{HTML}{FF2052}

\newcommand{\araa}{Annu. Rev. Astron. Astrophys.}   
\newcommand{\apj}{Astrophys. J.}   
\newcommand{\apjl}{Astrophys. J. Lett.}   
\newcommand{\aap}{Astron. Astrophys.}   
\newcommand{\aapr}{Astron. Astrophys. Rev.}   
\newcommand{\mnras}{Mon. Not. R. Astron. Soc.}   
\newcommand{\pre}{Phys. Rev. E}   
\newcommand{\planss}{Planet. Space Sci.}   
\newcommand{\pasj}{Publ. Astron. Soc. Jpn}   

\newcommand{\NO}{{\sf NO}}
\newcommand{\BO}{{\sf BO}}
\newcommand{\NL}{{\sf NL}}
\newcommand{\BL}{{\sf BL}}
\newcommand{\NH}{{\sf NM}}
\newcommand{\BH}{{\sf BM}}

\newcommand\pc{\, {\rm pc}}
\newcommand\kpc{\, {\rm kpc}}
\newcommand\Myr{\, {\rm Myr}}

\newcommand\Msun{\ensuremath{\, \mathrm{M}_\odot}}
\newcommand\erg{\, {\rm erg}}
\newcommand\kms{\, {\rm km~s}^{-1}}

\newcommand\EST{E_{\rm th}}

\newcommand{\mathbfss}[1]{\textbf{\textsf{#1}}}
\newcommand{\vect}[1]{\textbf{\textsf{#1}}}

\newcommand\cmcube{~ {\rm cm^{-3}}}
\newcommand\cs{ c_{\rm s}}
\newcommand\cplocal{ c_{\rm p}}
\newcommand\cv{ c_{ v}}

\jyear{2022}%

\theoremstyle{thmstyleone}%
\theoremstyle{thmstyletwo}%

\theoremstyle{thmstylethree}%

\received{\today}
\revised{\today}
\accepted{\today}
\begin{document}

\title[Supernova dust destruction in turbulent ISM]{Supernova dust destruction in the magnetized turbulent ISM}


\author[1,4]{\fnm{Florian} \sur{Kirchschlager}}\email{florian.kirchschlager@ugent.be}
\equalcont{These authors contributed equally to this work.}

\author*[2]{\fnm{Lars} \sur{Mattsson}}\email{lars.mattsson@su.se}
\equalcont{These authors contributed equally to this work.}

\author[3,2,5]{\fnm{Frederick A.} \sur{Gent}}\email{frederick.gent@aalto.fi}
\equalcont{These authors contributed equally to this work.}

\affil[1]{\orgdiv{Physics and Astronomy},  \orgname{Ghent University},  \orgaddress{\street{Krijgslaan 281-S9},  \city{Ghent},  \postcode{9000},  \country{Belgium}}}

\affil[2]{\orgdiv{Nordita},  \orgname{KTH Royal Institute of Technology and Stockholm University},  \orgaddress{\street{Hannes Alfv\'ens v\"ag 12},  \city{Stockholm},  \postcode{SE-106},  \country{Sweden}}}

\affil[3]{\orgdiv{Astroinformatics,  Computer Science},  \orgname{Aalto University},  \orgaddress{\street{PO Box 15399},  \city{Espoo},  \postcode{FI-00076},  \country{Finland}}}

\affil[4]{\orgdiv{Physics and Astronomy},  \orgname{University College London},  \orgaddress{\street{Gower Street},  \city{London},  \postcode{WC1E 6BT},  \country{UK}}}

\affil[5]{\orgdiv{School of Mathematics,  Statistics and Physics},  \orgname{Newcastle University},  \orgaddress{\city{Newcastle},  \postcode{NE1 7RU},  \country{UK}}}
\abstract{
Dust in the interstellar medium (ISM) is critical to the absorption and
intensity of emission profiles used widely in astronomical observations,  and
necessary for star and planet formation.  Supernovae (SNe) both produce and
destroy ISM dust. In particular the destruction rate is difficult to assess.
Theory and prior simulations of dust processing by SNe in a uniform ISM predict
quite high rates of dust destruction,  potentially higher than the
supernova dust production rate in some cases. Here we show simulations
of supernova-induced dust processing with realistic ISM dynamics including
magnetic field effects {and d}emonstrate how ISM inhomogeneity and magnetic
fields inhibit dust destruction. Compared to the non-magnetic homogeneous case,
the dust mass destroyed within 1\, Myr per SNe is reduced by more than a factor
of two,  which can have a great impact on the ISM dust budget.}




\maketitle

%

\section*{Introduction}
The interstellar medium (ISM) - gas and dust filling galactic space between the
stars - is critical to galaxy evolution and,  in particular,  accumulating and
cycling heavier elements.  {The survival of dust grains is a matter of debate
in cosmic dust evolution studies.} Several decades of research have clearly
established \cite[see][among many]{Barlow78,  Draine79,  McKee89,  Draine90,
Jones94,  Jones96,  Slavin04,  Jones11,  Bocchio2014,  Slavin2015,
Lakicevic2015, Mattsson16} that a supernova (SN) shockwave provokes ion
sputtering,  which can efficiently destroy dust grains.  The canonical model
outlined by McKee \cite{McKee89} suggests that supernovae (SNe) can effectively cleanse dust
from an ISM volume of gas mass equivalent to about $1000\,\Msun$.  Alongside
sputtering of dust grains,  fragmentation \citep[as described in,  e.g.,
][]{Borkowski95, Jones96} via grain-grain collisions can accelerate the
destruction rate, increasing further the dust cleansing efficiency
\citep[][]{KMG21}.  However,  the net dust-destruction
rate depends on complex gas dynamics,  including magnetic fields and
electrically charged grains.

Cosmic dust consists mainly of silicates and carbonaceous material,  which
profoundly impacts astronomical observations \citep{Weingartner01}.  Over long
timescales the atmospheres of evolved stars and molecular clouds provide the
dominant channels of dust production.  SNe are recognised both as
intermittent producers and destroyers of cosmic dust.  While ample
observational evidence \citep[e.g., ][]{Matsuura11,  Gomez12b,  Wesson2015,
Bevan2017,  NiculescuDuvaz2022} suggests a high degree of dust condensation
occurs in SN remnants,  there is no clear consensus on how much dust mass a
typical SN shockwave may destroy due to ion sputtering or grain-grain
collisions.  Strong quantitative constraints on the efficiency of dust
destruction are,  however,  of fundamental importance to correctly model the
matter cycle \citep{Mattsson16,  Zhukovska08,  Valiante11}.  Furthermore,  dust
abundances in {some} high-redshift galaxies appear to exceed expectations,
given current understanding of metallicity constraints and expected rates of
dust destruction \citep[see,  e.g., ][]{Dwek07,  Mattsson11b,  Rowlands14b}.

Of several decisive factors determining the dust survival rate,  two are
essential. First,  the ambient ISM gas density determines the reach of an SN
blast wave and size of the affected portion of the ISM \citep{McKee89}.
Second, the level of shock heating and accumulation of gas in the remnant shell
are critical to the sputtering and grain-grain collision rates
\citep[][]{Nozawa06, Bocchio2014,  Slavin2015,  Martinez2019}. Also,
frictional forces between gas and dust,  and Lorentz forces on charged grains
determine how well gas and dust are coupled,  which in turn may determine grain
survival.

Treating gas and dust as dynamically separate fluids in an inhomogeneous,
magnetised ISM may be of fundamental importance to the study of SN induced dust
destruction.  Such a study,  including all relevant forces and processes,  has
never been done. In particular,  the effect of magnetic turbulence on the
dynamics and survival of charged dust is poorly understood.  In \citep{KMG21},
shattering due to grain-grain collisions was included to model dust processing
in a uniform or modestly perturbed ISM,  increasing the dust destruction
efficiency significantly. This exacerbates the disparity between theoretical
estimates of dust destruction and observed dust abundances in {many} starburst
galaxies \citep{Ichikawa1994, Sodroski1994, Hutton2015}.

Here we apply the same dust processing models to a turbulent
magnetohydrodynamic (MHD) multi-phase ISM simulation to investigate how this
affects conservation of dust abundances. Turbulence causes the dust to
decouple from the gas, most so for the larger grains. Turbulence reduces the
dust losses from an SN blast wave by around 10\% and when the Lorentz force
acting on the dust is included this increases up to 50\%. The decoupling of the
dust is impeded by the Lorentz force, which is important to the dust survival.

\section*{Results}\label{sec:results}

We continue an MHD simulation of a supernova-driven turbulent ISM, in which
the the small-scale dynamo has saturated to provide a realistic turbulent
magnetic field within a turbulent multi-phase ISM. In three scenarios we
explode a single remnant in a diffuse region, a region of moderate gas density
and a case in which there is no explosion.  We then take the same 2D slice of
each case and apply two models of dust processing for a duration of 1\,Myr,
with and without including the effect of the Lorentz force on the dust
evolution. We compare the dust destruction between all models and with a case
without magnetic fields or turbulence obtained from \citep{KMG21}. The models,
their labelling convention and some indicative results are listed in
Table~\ref{tab:runs}.

\begin{table}
 \centering
 \caption{{\bf List of model parameters and cumulative dust losses.}
The models to which dust processing is applied are listed and denoted by the
prefix {\sf B} with the Lorentz force included and {\sf N} without.  The MHD
simulation runs are denoted by the suffix {\sf O} without an SN explosion, {\sf
L} for an explosion in the low density region or {\sf M} for an explosion in a
moderate density region.  $n_{\rm gas, 0}$ indicates the typical gas number
density of the explosion epicentre at $t=0$. Lorentz indicates whether or not
the magnetic effects are included for the dust processing. The accumulated dust
losses at 200 kyr,  500 kyr and 1 Myr are listed for each model. The ISM model of
\cite{KMG21} is uniform.}
 \begin{tabular}{p{1.2cm} p{1.2cm} c c p{1.2cm} p{1.2cm} p{1.2cm} }
 \hline
 Model&$n_{\rm gas, 0}$       &Lorentz&SN event&200 kyr          &500 kyr          &1 Myr            \\
      &$[\textrm{\, cm}^{-3}]$&       &        &[$\Msun$]        &[$\Msun$]        &[$\Msun$]        \\\hline
 \NO  & $\cdots$~          & no    &    no  &$\phantom{0}1.18$&$\phantom{0}6.36$&$18.2$           \\
 \BO  & $\cdots$~          & yes   &    no  &$\phantom{0}0.26$&$\phantom{0}2.42$&$\phantom{0}8.14$\\
 \NL  &$0.03$             & no    &    yes &$11.1$           &$29.0$           & $57.3$          \\
 \BL  &$0.03$             & yes   &    yes &$\phantom{0}5.3$ &$13.8$           & $28.4$          \\
 \NH  &$0.7$              & no    &    yes &$30.0$           &$46.8$           & $64.9$          \\
 \BH  &$0.7$              & yes   &    yes &$18.4$           &$26.5$           & $37.0$          \\[0.2cm]
\cite{KMG21} &$1.0$& no    &    yes & $48.6$          &$65.0$           & $70.9$          \\\hline
 \end{tabular}
 \label{tab:runs}
 \end{table}

\subsection*{Evolution of the MHD model}

\begin{figure*}
  \centering
   \resizebox{1.0\hsize}{!}{
 \includegraphics[trim=4.8cm 9.9cm 4.6cm 8.7cm,  clip=true, page=1]{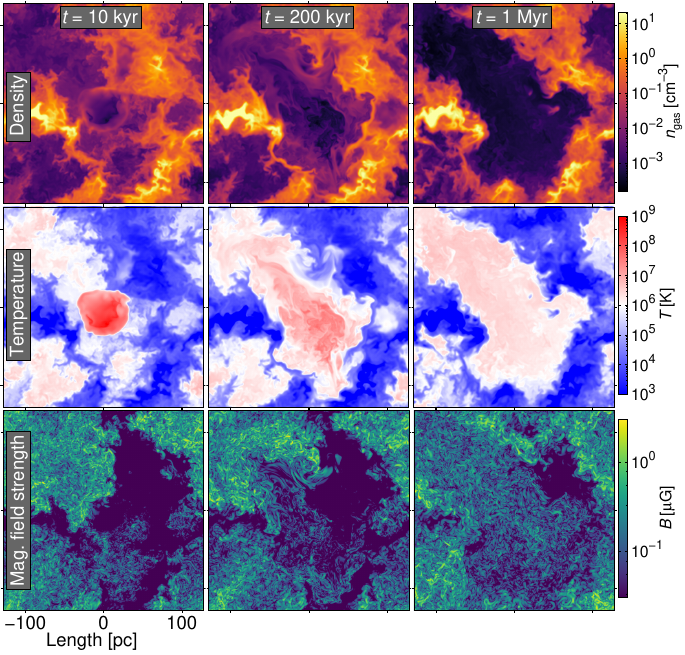}}
 \begin{picture}(0,0)
    \put(-345,294){\bf{\sf{a}}}
    \put(-345,194){\bf{\sf{b}}}
    \put(-345, 94){\bf{\sf{c}}}
  \end{picture}
	  \caption{
{\bf{Gas structure following explosion in low density region (\BL/\NL)}}.
{\bf a} Snapshots of the gas density at times $t = $ 10 kyr,  200 kyr,   1 Myr. {\bf b} Same snaphots of gas temperature.
{\bf c} Same snaphots of magnetic field strength.  The SN
explodes in a low density region located at the centre of each image.  A short
movie showing the evolution is presented
\href{here}{https://youtu.be/Mz2cuEVm_eY}
(movie frame rates $f = 133.3$ frames
s${}^{-1}$), from the time the SN explosion is introduced ($t=0$) until $1\,$Myr later.
A short movie of the turbulent ISM without SN explosion (model \BO/\NO) is available
\href{here}{https://youtu.be/HEJ-4CZl-ZQ} ($f = 33.3$ frames s${}^{-1}$).
The movies and data are publicly available with the \href{Source Data}{https://etsin.fairdata.fi/dataset/602bb9a6-0626-43db-9073-054bd3332fff/data}.
	  }
   \label{fig_gas_evolution_a}
  \end{figure*}

\begin{figure*}
  \centering
   \resizebox{1.0\hsize}{!}{
 \includegraphics[trim=4.8cm 9.9cm 4.6cm 8.7cm,  clip=true, page=1]{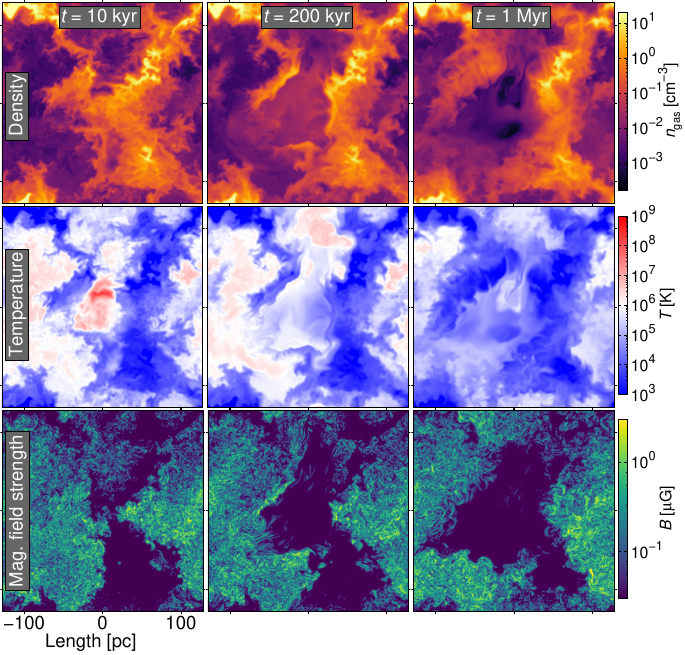}}
 \begin{picture}(0,0)
    \put(-175,307){\bf{\sf{a}}}
    \put(-175,207){\bf{\sf{b}}}
    \put(-175, 107){\bf{\sf{c}}}
  \end{picture}
	  \caption{
{\bf{Gas structure following explosion in moderate density region (\BH/\NH)}}.
{\bf a} Snapshots of the gas density at times $t = $ 10 kyr,  200 kyr,   1 Myr. {\bf b} Same snaphots of gas temperature.
{\bf c} Same snaphots of magnetic field strength.  The SN
explodes in a moderate density region located at the centre of each image.  A short
movie showing the evolution is presented
\href{here}{https://youtu.be/oGXh6piIqlA} (movie frame rates $f = 133.3$ frames
s${}^{-1}$), from the time the SN explosion is introduced ($t=0$) until $1\,$Myr later.
The movies and data are publicly available with the \href{Source Data}{https://etsin.fairdata.fi/dataset/602bb9a6-0626-43db-9073-054bd3332fff/data}.
	  }
   \label{fig_gas_evolution_b}
  \end{figure*}

Fig.~\ref{fig_gas_evolution_a}{\,{a}} displays slices depicting evolution
of the gas density from a simulation of SN-driven turbulence,  containing the
remnant of an explosion located in low ambient density (\NL\ or \BL) and
Fig.~\ref{fig_gas_evolution_b}\,{a} an explosion located in moderate
density ambient ISM (\NH\ or \BH). The models are denoted by the prefix B with the Lorentz force included and N without.
The remnant retains a signature of its
self-similar spherical origins at 10 kyr,  but subsequent expansion is
irregular due to the multi-phase ISM structure.  Prior to explosion slices for
all models are identical,  only shifted vertically to locate suitable ambient
density at each explosion epicentre.  Models omitting the SN explosion are
evolved separately  (models \NO\ or \BO) to measure background processing
of dust, processing which occurs due to the background turbulence where the SN blast wave has not reached. Their profiles barely alter from those already visible outside the
initial remnant regions at 10\,kyr in Figs.~\ref{fig_gas_evolution_a} and
\ref{fig_gas_evolution_b} over 1\,Myr. Maps at additional time-steps for the
gas impacted by the blast waves are presented in Supplementary~Fig.~1 and 2.

Corresponding temperatures are illustrated in
Fig.~\ref{fig_gas_evolution_a}\,{b} and \ref{fig_gas_evolution_b}\,{b}.
High radiative losses from cooling in the dense medium  (Fig.~\ref{fig_gas_evolution_b}) reduces the
strength of the blast wave early on. In this model reflected shock waves are
weaker and damped more rapidly by its relatively higher density remnant
interior.  The blast wave in the diffuse gas (Fig.~\ref{fig_gas_evolution_a}) is initially much faster
and shows stronger reflected shock waves as the blast wave later encounters
dense regions and propagates at high velocity in the relatively diffuse remnant
interior.

Fig.~\ref{fig_gas_evolution_a}\,{c} and \ref{fig_gas_evolution_b}\,{c} display magnetic field strength,
which has been amplified by dynamo action of SN-driven turbulence from a
sub-nanoGauss random seed field.   By 1\, Myr the magnetic field grows
substantially inside the diffuse remnant,  evidence of turbulent dynamo in the
hot gas \cite[see][]{GMKS22},  while in the dense remnant magnetic field is
mostly evacuated with the blast wave or dissipated. The varying location and
topology of the magnetic field during the evolution of each remnant may impact
the dust processing.

Model \NO\ or \BO\ without new SNe continue to evolve the turbulence relatively
slowly,  the density and magnetic field profiles remaining similar to their
depiction at 10~kyr in Fig.~\ref{fig_gas_evolution_b}\,{a} and c.
Velocities in the hot gas are of order 100~km~s$^{-1}$ and in warm gas
10~km~s$^{-1}$.  Ambient ISM background dust processing is,  therefore,  likely
to be significant and complex,  in contrast to zero processing in a uniform ISM
\citep[][]{Hu2019, Slavin2015, Martinez2019,  KMG21}.  Some background
processing might be discounted from the total dust losses in the  SNe models.

\newpage

\subsection*{Dust destruction effect of magnetic fields}

\begin{figure*}
   \resizebox{1.05\hsize}{!}{\hspace*{-0.3cm}
 \includegraphics[trim=0cm 0.5cm 8.1cm 0.4cm,  clip=true, page=1]{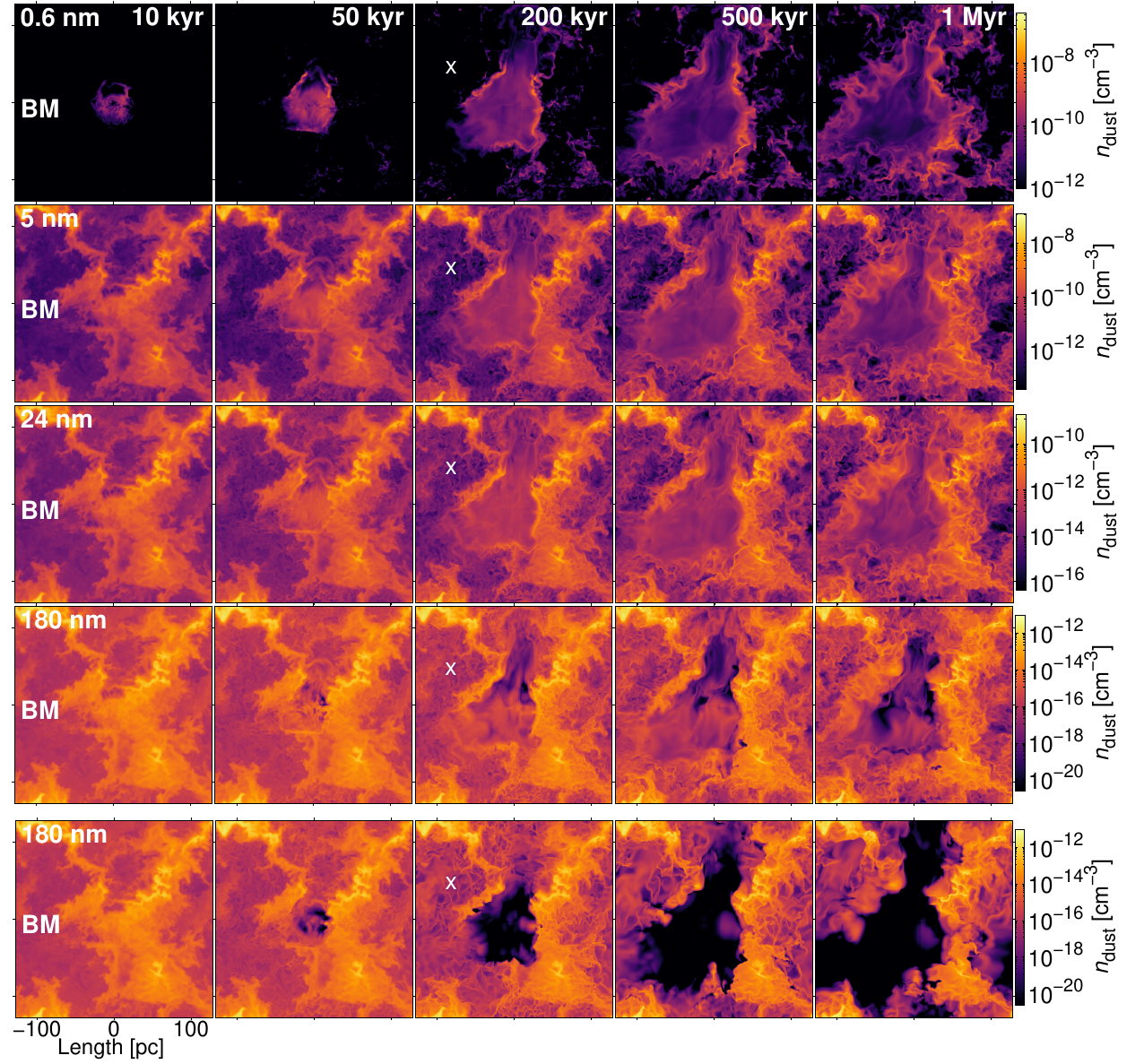}}
 \begin{picture}(0,0)
    \put(-10,334){\bf{\sf{a}}}
    \put(-10,267){\bf{\sf{b}}}
    \put(-10,205){\bf{\sf{c}}}
    \put(-10,140){\bf{\sf{d}}}
    \put(-10, 75){\bf{\sf{e}}}
  \end{picture}
   \caption{{\bf{Dust distribution for various grain sizes.}} Snapshots of dust
density following the SN explosion in the moderate density region.  {\bf{a}} --
{\bf{d}} The distribution of 0.6,  5,  24,  180$\, $nm grains, respectively for
model \BH, which includes the effect of Lorentz force on the dust; the colour
scale is fixed for each row. {\bf{e}} Distribution of 180$\, $nm grains for
model \NH, without Lorentz force effects. The white ‘X’ identifies a region at
which the Lorentz force reduces the turbulent scale of the dust (see text for
details). A short movie
showing the temporal evolution (model \BH; movie frame rate $f = 133.3$ frames
s${}^{-1}$) is available \href{here}{https://youtu.be/w8ZJqZK63KY}. For comparison,  a short movie showing the
explosion in the diffuse ISM (model \BL; $f = 133.3$ frames s${}^{-1}$) or the
temporal evolution without SN explosion (model \BO; $f = 33.3$ frames
s${}^{-1}$) is available \href{here}{https://youtu.be/byw8LQ38i8M}
and \href{here}{https://youtu.be/36nB1aAL-2o}, respectively.
The movies and data are publicly available with the \href{Source Data}{https://etsin.fairdata.fi/dataset/602bb9a6-0626-43db-9073-054bd3332fff/data}.
}   \label{fig_highdensity_mag_dust}
  \end{figure*}
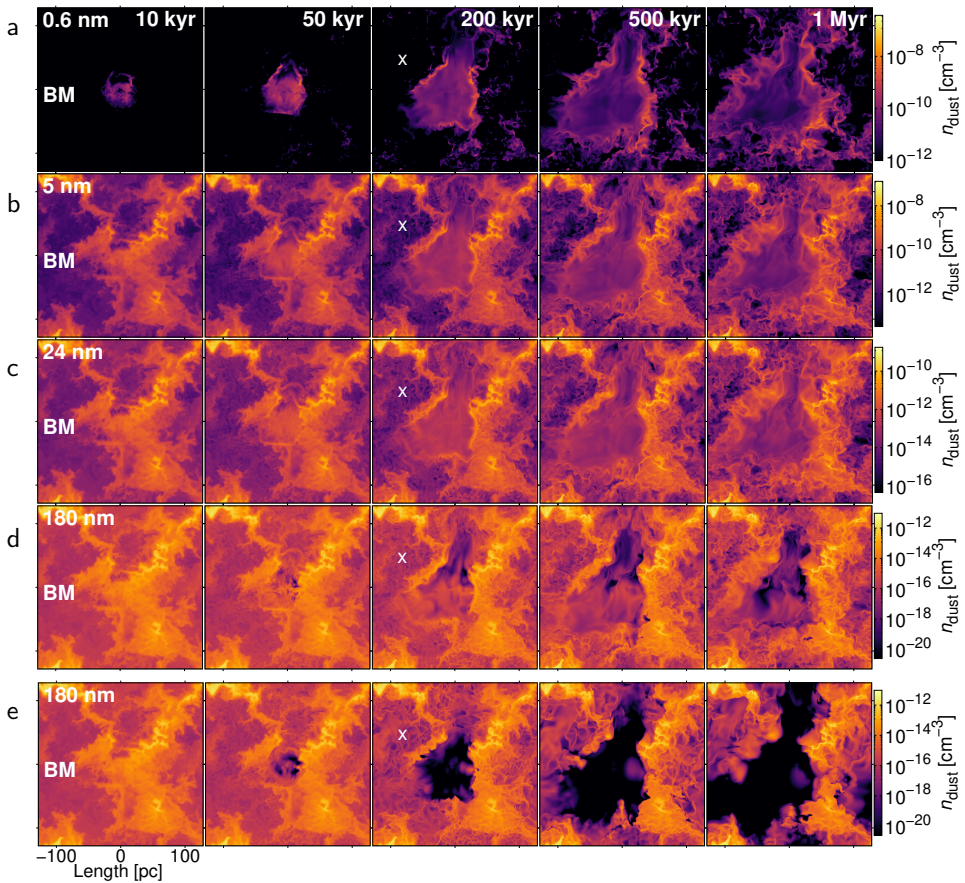

In Fig.~\ref{fig_highdensity_mag_dust} we illustrate the evolution of dust of
different grain sizes for model~\BH\ (see Table~\ref{tab:runs}).  Panels {a} -- {d}
 depict dust in bins of increasing grain size. The smallest grain
size ({a}; 0.6\, nm) is not included in the initial distribution,  but results from
shocks and turbulence fragmenting and sputtering larger grains.  Beyond 200\,
kyr,  larger dust grains in the remnant interior are lost or advected to the
shell at a rate increasing with grain size.  For the same explosion where the
Lorentz force is neglected in processing the dust (model~\NH),  the
annihilation of dust in the remnant interior is far more efficient.
Fig.~\ref{fig_highdensity_mag_dust}{\,e} shows
this comparison relative to Fig.~\ref{fig_highdensity_mag_dust}{\,d} for the 180\, nm bin grain size.  Dust destruction is
significantly reduced by the presence of a magnetic field. The dust maps of the
SN explosions in the low and moderate density regions are presented in
Supplementary~Fig.~3 and~4.

Due to ambient turbulence the dust processing continues outside the remnant.
Such a region is identified by a white {\sf `X'} in
Fig.~\ref{fig_highdensity_mag_dust}. At 10\, kyr the dust in these regions has
a smooth distribution,  scaling directly with the gas. By 200\, kyr the dust
decouples from the gas to form filamentary structure.  Fourth and fifth row
comparison reveals that the Lorentz force reduces the filamentary scales and
contrast in the dust.  The gas velocity is the same in both models,  indicating
that the Lorentz force on the dust inhibits it decoupling from the gas. Inside
the remnant the large grains survive better with the Lorentz force than
without.

\begin{figure*}
 \resizebox{\hsize}{!}{
 \includegraphics[trim=1.4cm 0.2cm 0.9cm 0.4cm,  clip=true, page=1, height = 4cm]{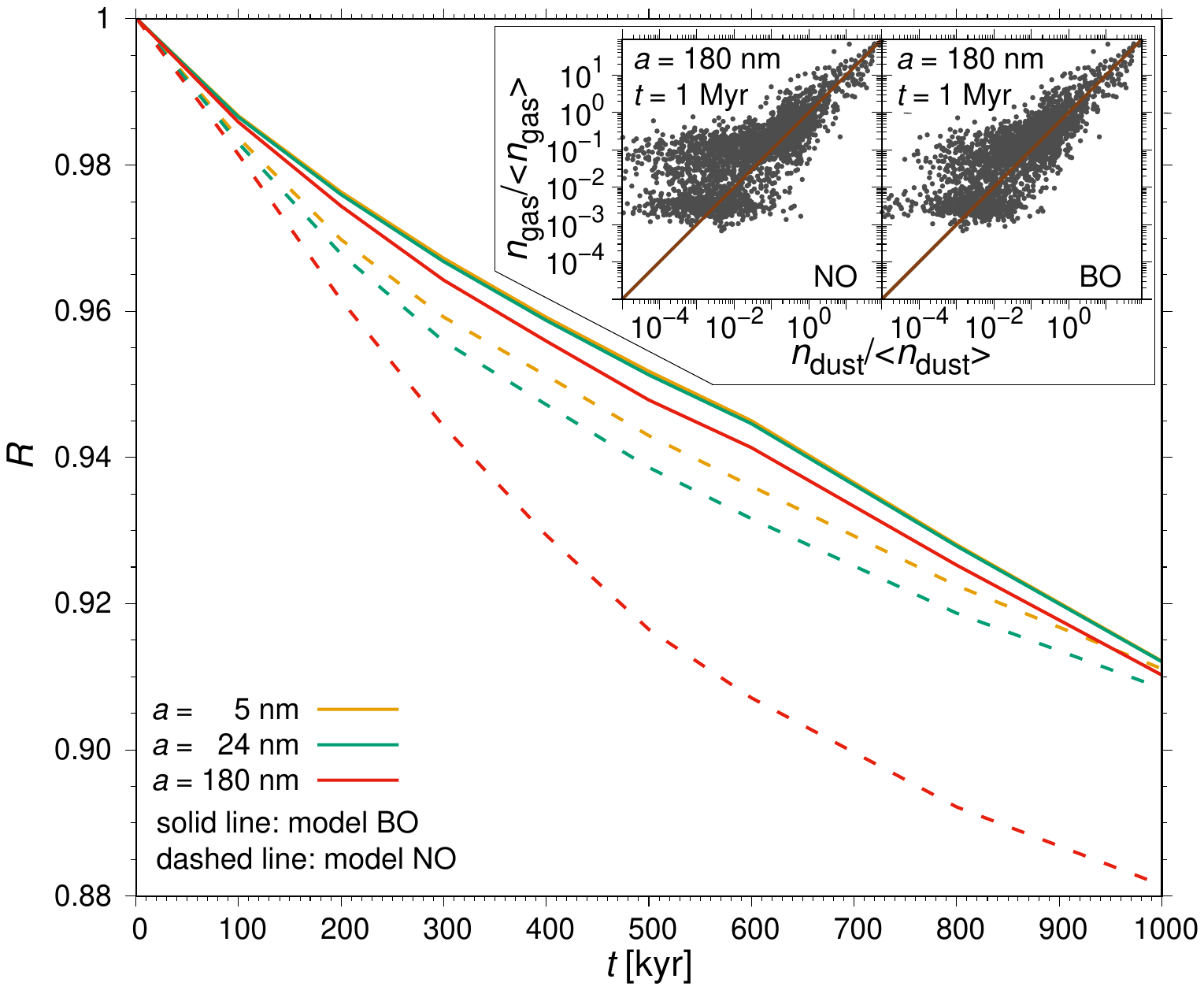}
 }
  \caption{{\bf{Reduction of correlation between dust and gas.}} Variation over time of the Pearson
correlation coefficient $R$ due to the background dust processing between
the logarithmic gas density and the logarithmic dust density
for models \BO\ (solid lines) and \NO\ (dashed lines) at
different grain sizes (different colors). The inset shows the scatter plots
of normalized gas and dust density for the grain size $180\, $nm at $1\,
$Myr, with the brown line indicating exact correlation.
Source data are provided as a Source Data file.}
  \label{fig_scatter}
  \end{figure*}

Fig.~\ref{fig_scatter} shows for sample grain sizes the Pearson
correlation coefficient $R$ for the correlation between the logarithmic gas and
dust number densities.  Initially $R=1$,  but the correlation decreases over
time. Dust-gas coupling is greater when the Lorentz force is present,  more so
the smaller the grain size \citep[see also][]{Mattsson18a, HBSM22}.
Consequently,  the weakest coupling occurs for the largest grains without
Lorentz force after 1$\, $Myr (see inset scatter plots in
Fig.~\ref{fig_scatter} and Supplementary Figs.~5 and 6).

\begin{figure}
 \resizebox{\hsize}{!}{
 \includegraphics[trim=2.0cm 1.5cm 1.3cm 2.1cm,  clip=true, page=1, height = 4.5cm]{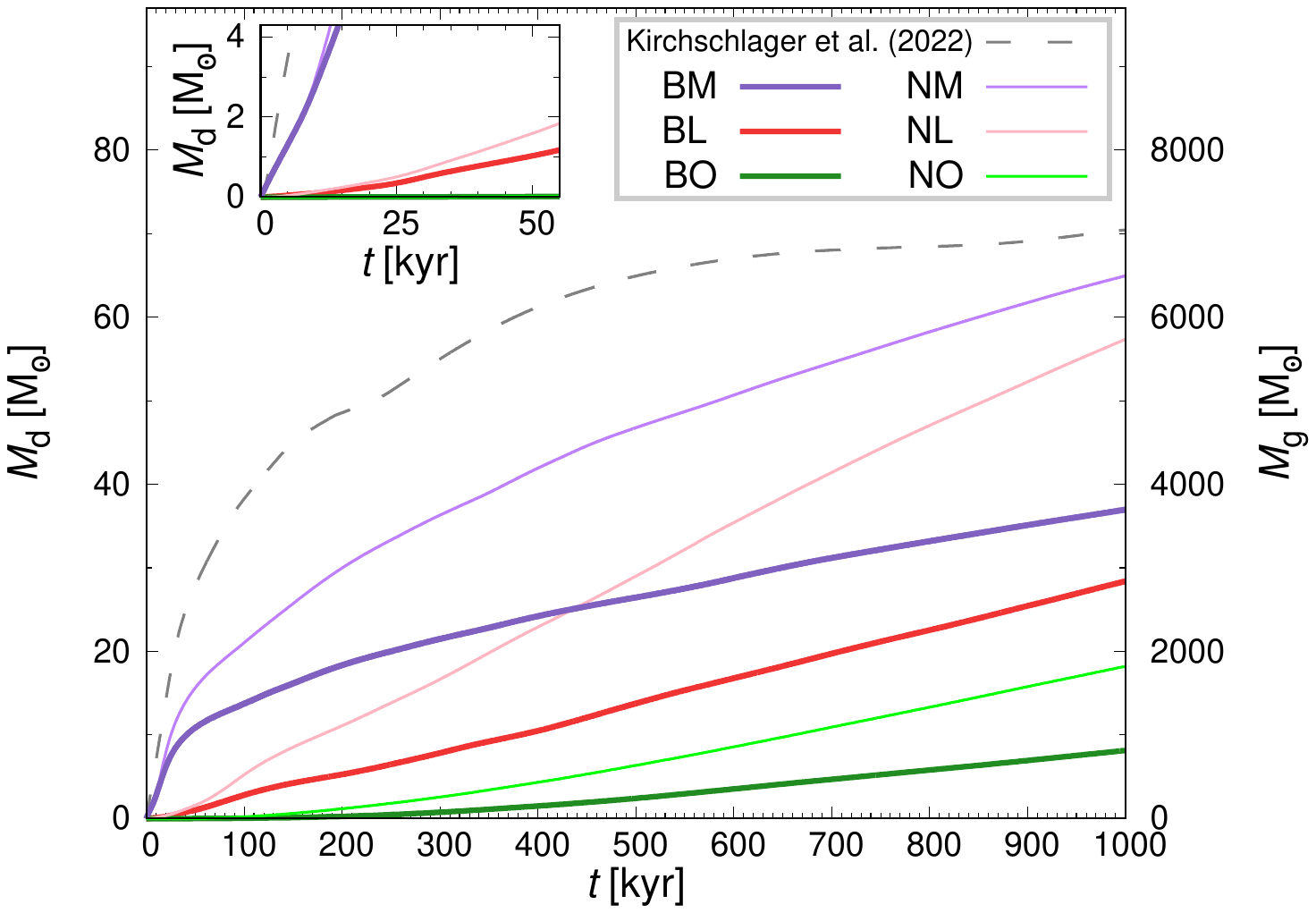}
 }
  \caption{{\bf{Cumulative dust mass destroyed and cleared gas mass.}} Total
dust mass destroyed $M_{\rm d}$ and cleared gas mass $M_{\rm g}$ as a function
of time,  for each model as listed in the legend. The scales on each $y$-axis
are directly equivalent,  with $M_{\rm g}=100 M_{\rm d}$. The inset shows the
dust mass destroyed within the first 50\, kyr.
Source data are provided as a Source Data file.}
  \label{fig_dustmass_evolution}
\end{figure}

In Fig.~\ref{fig_dustmass_evolution} we display the cumulative mass of dust
destroyed in each model over the first Myr.  For all three cases more dust is
destroyed when the Lorentz force is excluded.  Background processing only
(green) is negligible up to 100\, kyr,  after which differences due to the
Lorentz force become significant,  such that within 1\, Myr dust losses are
reduced by more than half.  The rate of dust destruction increases until
$200-300$\, kyr,  becoming steady thereafter with rates of approximately
$23\, \Msun\Myr^{-1}$ and $11\, \Msun\Myr^{-1}$ for \NO\ and \BO, respectively.
This suggests an approximate 300\, kyr transient redistribution of the dust
from its initial gas-coupled condition into a statistical steady state,  with
dust depletion settling at 23 or $11\, \Msun\Myr^{-1}$,  respectively. Beside
the total mass of destroyed dust,  the dust destruction fraction  of the entire
domain characterizes the dust processing (see Supplementary Fig.~8). For all
models presented in Table~\ref{tab:runs},  the fractions are less than 3\,
per~cent.

Where explosions are sited in low (red) or moderate (purple) density gas the
dust destruction is initially not very sensitive to Lorentz force effects.
Magnetic effects significantly reduce dust destruction after $20-50\, $kyr.

\subsection*{Effect of SN ambient ISM gas density on dust destruction}

The first 50\, kyr are critical to the effect of ambient gas density at the SN
epicentre on total dust destruction.  Less than $2\, \Msun$ of dust has been
destroyed in models \NL\ and \BL\ (Fig.~\ref{fig_dustmass_evolution}; red),
compared to  over $10\, \Msun$ and $15\, \Msun$  in models \NH\ and \BH\
(purple),  respectively,  or $28\, \Msun$ for \citep{KMG21}.

Subsequently the dust destruction rates slow considerably up to about 300\,
kyr.  For moderate ambient density it continues to slow,  while otherwise it
increases slightly.  After 400\, kyr the rate of destruction becomes steady up
to 1\, Myr: around $57\Msun\Myr^{-1}$ for \NL\ (light-red); $30\Msun\Myr^{-1}$
for \BL\ (dark-red); $38\Msun\Myr^{-1}$ for \NH\ (light-purple); and
$22\Msun\Myr^{-1}$ for \BH\ (dark-red).

The dust destruction rate is highest soon after the SN explosion,  with losses
higher where the ISM is dense. After about 500\, kyr,  the blast wave from a
dense epicentre loses speed and dust losses slow as it sweeps through more
diffuse regions. From a diffuse epicentre the rate slightly increases later as
more dense regions are affected.

In all cases the dust destroyed is lower than in a uniform ambient gas density
of $1\textrm{\, cm}^{-3}$ without Lorentz force (\citep{KMG21}; gray-dashed line in
Fig.~\ref{fig_dustmass_evolution}).  The dust destruction rate is higher in
\citep{KMG21} than in other models until around 500\, kyr,  after which it becomes
negligible.

\subsection*{Impact of dust processing on the dust distributions}

\begin{figure*}
 \resizebox{\hsize}{!}{\hspace*{-0.6cm}
 \includegraphics[trim=1.9cm 0.6cm 0.35cm 0.5cm,  clip=true, page=1]{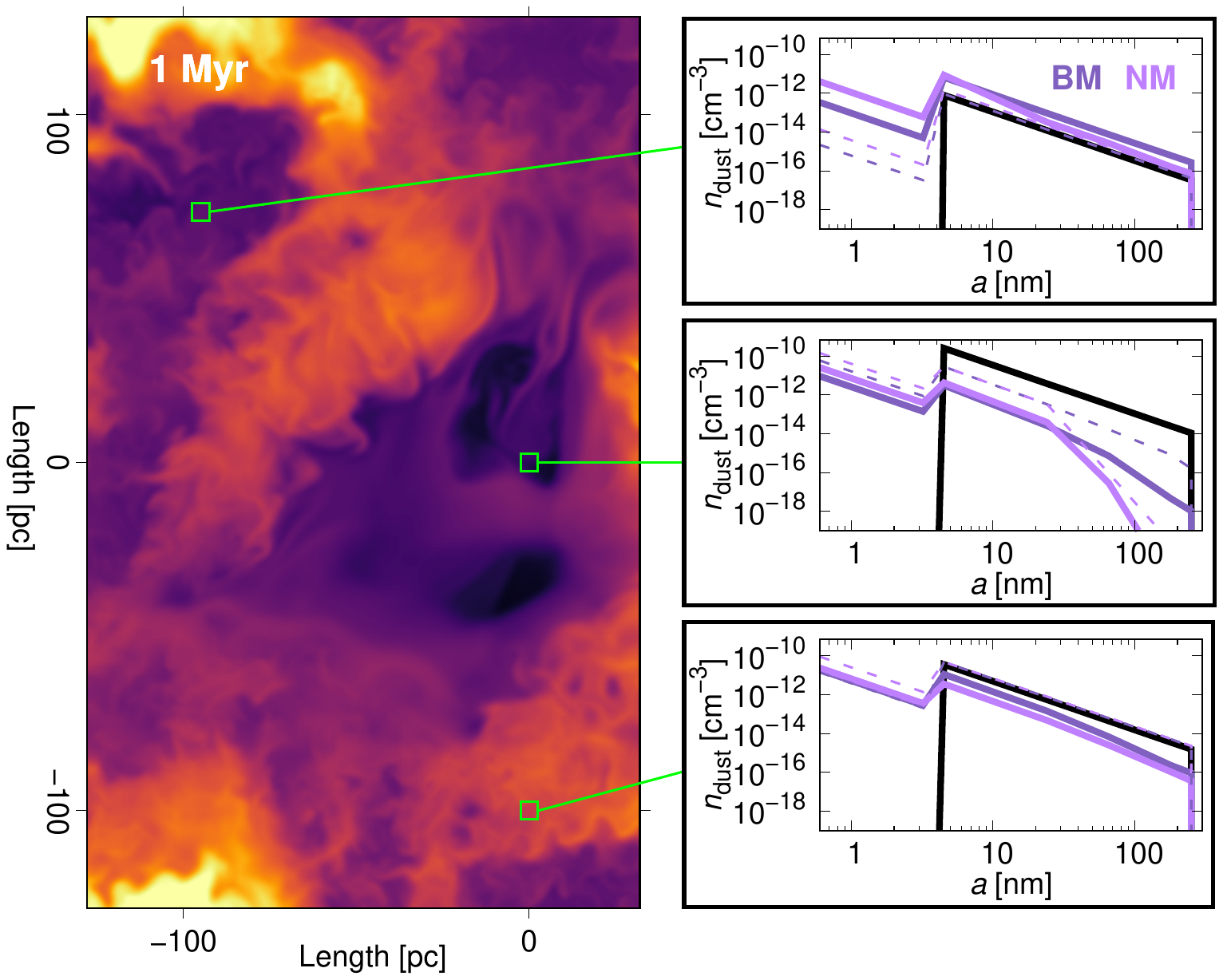}
 }
 \begin{picture}(0,0)
    \put( -5,270){\bf{\sf{a}}}
    \put(177.5,270){\bf{\sf{b}}}
    \put(177.5,185){\bf{\sf{c}}}
    \put(177.5, 98){\bf{\sf{d}}}
  \end{picture}
  \caption{
{\bf{Regional variation of dust distribution at various times.}}
{\bf{a}} Location within the domain with an explosion in moderate ambient
gas density of three specimen regions (green boxes) representing
two unshocked  regions  (top and bottom) and the shocked region (center).
The green boxes have a size of $5\, \text{pc}\times5\, $pc.  {\bf{b}} The
dust density initial MRN distribution (black), distribution after 200\, kyr
(dashed lines) and after 1\, Myr (solid lines) for an unshocked region of low
gas density, with Lorentz force (dark-purple; model \BH) and without Lorentz
force (light-purple; model \NH).  {\bf{c}} Dust density distribution for the
central shocked region. {\bf{d}} Dust density distribution for an unshocked
region of moderate gas density.  The dust distributions show the average
density for each grain size in each specimen.  In the shocked region,  a
significant amount of dust grains is removed or destroyed by the shock while
the change in the unshocked regions is due to background processing
(transport and destruction).
Source data are provided as a Source Data file.
}
  \label{fig_gsd}
  \end{figure*}

In Fig.~\ref{fig_gsd} the initial,  intermediate,  and final dust number
densities for the models \NH\ and \BH\ are shown in panels b -- d for selected regions identified in panel a.  The
dust density distributions by grain size $a$ are the averages for each box, 5\,
pc $\times$ 5\, pc square,  with insets showing initial MRN
\cite{Mathis77} (black solid line), 200\, kyr (dashed lines) and 1\, Myr
(solid lines) profiles.  Initial dust densities,  following a fixed power-law
spanning 5 -- 250 nm,  scale with local gas density and differ between regions.

The region at the explosion epicentre represents more dense ISM impacted
very early by the blast wave. Therefore,  the dust densities (Fig.~\ref{fig_gsd}\,c) show significant
alteration.  After 1 Myr,  the dust densities for all pre-existing grain sizes
reduce about two orders of magnitude -- either swept or destroyed by the blast
wave (through sputtering,  fragmentation or vaporization). Destruction is
higher for large grains,  especially when the Lorentz force is neglected (\NH).
With the Lorentz force,  better dust-gas coupling reduces drag forces, relative
motions between gas and dust,  and between grains of differing size.  On the
other hand,  the Lorentz force enables a small but significant proportion of
dust to get behind the shock wave,  including large grains
(Fig.~\ref{fig_highdensity_mag_dust}{\,d} -- {e}). Most of the
destruction in this central region occurs already within 200\, kyr.

Grain shattering produces a power-law distribution of smaller fragments,
including a {fragmental} range below the short end of the initial size
distribution.  Destruction of pre-existing grains are lower when the Lorentz
force is considered,  so final number densities at all fragment sizes are
lower.  Larger fragments exceeding 5\, nm add to number densities at the short
end of the initial size distribution.  The outcome is a distribution well
approximated by two power-laws: fragments and sputtered grains comprising the
fragmental distribution and the modified initial grain size distribution with
radii above 5 nm.

While fragmentation changes the shape of the grain-size density distribution,
the actual dust mass destroyed by shattering is rather small.  However,
vaporization and sputtering can destroy dust mass,  the latter especially
effective for small dust grains. Fragmentation and sputtering are thus
synergistic and work together by first shattering larger grains into smaller
pieces and subsequently destroying dust mass by sputtering of the fragments
\citep{Kirchschlager2019}.

The two other regions in Fig.~\ref{fig_gsd}\,a are located in
unshocked gas beyond the 1\, Myr reach of the blast wave.  Alterations in the
dust density distribution are caused by transport and destruction due to gas
dynamics only.  The unshocked region with moderate gas density
(Fig.~\ref{fig_gsd}\,d) shows a reduction of dust densities, but at a
significantly lower rate than in Fig.~\ref{fig_gsd}{\,c}. The reduction
in dust densities is again larger for the model \NH\ and can be assigned to
dust destruction or the flow of gas from one cell to another.  This gas flow
is particularly crucial for the lowest density region
(Fig.~\ref{fig_gsd}\,b),  where the amount of dust is even increased
compared to the initial conditions. The increase of gas density is 21\, per
cent over the Myr.  For both unshocked regions  little change to the
initial dust distribution occurs within the first 200\, kyr (dashed lines in
Fig.~\ref{fig_gsd}\,b and d).

In the lower box  without the Lorentz force,  higher numbers of fragmental
grains within 200\, kyr result from faster fragmentation of larger grains,
reducing by 1\, Myr to the same levels applying with Lorentz force.  Given
fragmentation without Lorentz force is higher at the beginning,  small grain
losses due to sputtering must reduce when the Lorentz force is considered.

In contrast to the other boxes,  the number densities of the fragmental
distribution in the upper box continue growing after 200\, kyr.  Much of this
follows gas inflows,  with which small grains are even better coupled.
However,  the ratio of fragmental densities to the 5\, nm grain number
density is smaller than in the other boxes,  more so when the Lorentz force
is considered,  which must mainly be explained by low fragmentation of larger
grains rather than high sputtering of small grains.  As two boxes are
unshocked,  the alteration of dust densities indicates the importance of ISM
inhomogeneity and background turbulence.

\section*{Discussion}\label{sec:conclusion}\label{sec:discussion}

We have conducted the high resolution MHD simulations that explicitly follow dust
destruction by the combined effects of grain-grain collisions and sputtering
of an SN blast wave in a turbulent multiphase,  magnetized ISM. Several factors
affect dust processing induced by an SN,  but also the background processing
due to ISM dynamics is considerable (see green profiles in
Fig.~\ref{fig_dustmass_evolution}). The background processing rate outside the
SN shock front is $11\Msun\Myr^{-1}$ which is equivalent to
$650\Msun\Myr^{-1}\kpc^{-3}$. The ISM density variability created by turbulence
can enhance the dust processing rate \citep[see e.g., ][]{Hirashita09,
Hirashita10, Mattsson20, Mattsson20b},  but an inhomogeneous ISM also appears
to protect dust from propagating SN shocks.

 In regions of moderate mean ambient density $\langle n_{\rm gas} \rangle \approx 1
\textrm{\, cm}^{-3}$,  a considerable amount of dust near the explosion
epicentre is lost early in the blast wave.  Stars associated with OB clusters
may explode in regions of higher ambient densities ($\gg 1\textrm{\,
cm}^{-3}$). How commonly this occurs, or whether they evacuate the ambient
medium in which subsequent explosions occur,  is unclear.  {O}nly 15\% of SNe
surveyed {\cite{HY09} interract with} the small fractional volume of the ISM
that comprise high density molecular clouds {of order $10^5\textrm{\,
cm}^{-3}$}.  Such observational signatures of interaction with dense gas can
even arise later in the life of a remnant.  {On the other hand,  such emissions
from cloud densities below $10^5\textrm{\, cm}^{-3}$ are difficult to detect,
so the SN rate in dense regions may be underestimated \cite{Sofue20}.}
Concluding the separation of clouds from their stellar progeny occur mainly
ahead of the SNe,  \cite{Gatto15} find it likely that SNe will more often occur
in ambient diffuse ISM,  {subject to their limited resolution and absent
ionization}.  To examine processing in much higher density locations,  or
earlier in remnant evolution would require resources and inclusion of
additional physics beyond the scope of this paper.

{Nevertheless,  let us consider what to expect of dust destruction at high gas
densities.  Contrary trends are subject to the actual density,  its spatial
extent and the shock velocity.  If the ambient mass is sufficiently high the
shock will quickly dissipate before it can penetrate the entire region,
resulting in almost no dust processing in the outer regions.  Therefore,  the
dust survival rate in the high density region could be arbitrarily large if the
region is sufficiently massive.  On the other hand,  where the mass is
insufficient the shock would overrun the entire region and process all the
material.  The expected outcome still depends on the extent to which
self-shielding effects due to higher densities compensate the higher
destruction due to greater dust abundances.  In areas that are closest to the
explosion centre,  the destroyed dust mass will be increased due to higher
sputtering and grain-grain collision rates,  which is confirmed by the early
dust evolution at moderate densities in this study and also by simulations of
ejecta clumps that are overrun by shocks (e.g. \cite{Kirchschlager2019}).  In
addition to adiabatic effects,  radiative gas cooling is accelerated in high
density regions and consumes a substantial quantity of the energy available to
the shocked gas,  which in turn is no longer available for dust destruction
processes.  In summary,  the balance between these conflicting effects is
difficult to estimate and we shall require simulations to include sufficiently
high density regions.}

The plausibility of the destroyed dust masses derived in \citep{KMG21}  has been checked
(Section 5.1 in \citep{KMG21}) against the results of previous studies \cite{Slavin2015,
Martinez2019,  Hu2019}. Deviations can be mainly retraced to different
conditions (explosion energies,  magnetic field strengths,  gas-to-dust-mass
ratios),  neglecting physical processes (grain-grain collisions,  kinetic
sputtering),  or different evolution times. In the present study,  the higher
density inhomogeneity induced by SN-driven turbulence reduces dust losses by 6
-- 14~$\Msun$ over $1\Myr$ relative to the uniform ambient ISM model of \citep{KMG21}.
However,  it is perhaps more indicative to consider the results over only 500\,
kyr,  within which timescale neighbouring SNe might be expected to interact
with the remnant. Thus far the multiphase inhomogeneous ISM reduces the dust
destruction compared to \citep{KMG21} by 18 -- 36~$\Msun$.

While the authors anticipated that including the effects of the Lorentz force
might affect dust destruction,  its impact is surprisingly strong.  When
including the Lorentz force acting on charged grains within 1\, Myr about 28 --
29~$\Msun$ of dust is additionally conserved in the multiphase ISM than without
Lorentz forces. Within 500 kyr the total destroyed dust mass is $13.8$
($26.5$)\, $\Msun$ for the low (moderate) density explosion site. Lorentz forces
on charged dust reduce destructive grain-grain interaction (in particular
fragmentation of large grains),  which lowers the overall dust-destruction
rate. Fewer small grains are produced.  The magnetic field at least halves the
dust losses due to background processing,  but has even more impact against SN
shocks. The lower dust destruction in the ISM at higher magnetic fields also
confirms results of previous studies \cite[e.g., ][]{Slavin2015}.

Due to absence of both galactocentric differential rotation and stratification,
here large-scale dynamo is not present. The magnetic field generated purely by
a small-scale dynamo has only a turbulent structure and saturates at a strength
an order of magnitude weaker than might be expected in disk galaxies
\citep{SSFBK15, Federrath16, GMKS21, GMKS22}.  The large-scale dynamo adds a
strong field ordered along the plane of the disc \citep{Gressel08b,
Gent:2013a},  but also entrains a turbulent field over ten times stronger than
obtained here \citep{GMK23}. It is likely that this stronger turbulent
component would protect dust even more effectively.  Large-scale fields are
observed to be weaker than the turbulent component \citep{Beck15},  so are
unlikely to undermine,  and may even enhance,  the effect of the turbulent
field.

In these models,  we assume an initial dust abundance proportional to the gas
density (initial gas-to-dust-mass ratio 100). It is not our aim to study the
early Universe or pristine ISM,  so a much higher gas-to-dust-mass ratio can be
ignored here. We find from all models for regions with only background
processing that the dust tends to aggregate in more filamentary structures than
the gas. A turbulent magnetic field appears to cluster the dust on smaller
scales and with reduced filamentary structure. Charged grains seem to be better
protected from destruction in the presence of a magnetic field.  We hypothesize
that this is due to the fact that the Larmor time is affected by the variation
of the local Alfv\'enic Mach number,  which means the magnetic force on the
dust alters dust dynamics and hence the clustering of dust grains.

Dust survival in the ISM as an SN shock wave propagates through it can vary due
to mainly three factors: the distribution (inhomogeneity) of the ISM gas,  the
mean density of the gas and the ambient dust abundances at the site of the SN.
The importance of the latter two factors were quite expected.  The first has a
surprisingly strong effect.  When factoring in the effect of the Lorentz force
from a turbulent magnetic field these have a net result stronger than we
anticipated. Overall,  the three factors listed above reduce dust losses
compared to the homogeneous case of the reference model (\citep{KMG21}) without the
Lorentz force by between 28\% and 55\% within 500\, kyr.  Including the effects
of the turbulent magnetic field reduces this even further,  overall by between
60\% and 79\% (see Table~\ref{tab:runs}).

{A destroyed dust mass of around $30\, \text{M}_\odot$ (including
turbulence and magnetic fields) is higher than found in other studies,  as e.g.
in \cite{Slavin2015} (about $10\, \text{M}_\odot$).  However,  we can reconcile
the deviation by accounting for the larger explosion energy ($1\times 10^{51}\,
$ergs vs $0.5 \times 10^{51}\, $ergs),  lower gas-to-dust-mass ratio (100 vs
163) and longer evolution times ($1\, $Myr vs $540\, $kyr).  Although including
turbulence and magnetic fields reduces the destroyed dust mass by a factor
about 2,  the total amount of destroyed dust is still very high. For
conditions as in \cite{Slavin2015} this would result in destroyed dust masses
of approximately $5\, \text{M}_\odot$. Assuming that a single core-collapse SN produces
dust of order 1 solar mass,  it is significantly less than the dust
destruction,  and thus a net dust destroyer. Including magnetic fields and
turbulence signficantly reduces the burdon for dust production sources,  namely
AGB stars (not in the early Universe) or the ISM \cite{Jenkins09,  Dwek16},
necessary to account for levels of net dust in the universe. Further effects
that could potentially reduce the dust destruction,  at least in the first few
kyr,  e.g. stellar wind-blown bubbles and shells \cite{Martinez2019} or
large-scale dynamos,  are promising and have to be considered in future
studies.}

\section*{Methods}\label{sec:methods}

\subsection*{Numerical simulations}\label{sec:mhd_methods}

For the ambient turbulent ISM,  within which we process the dust,  as an
initial condition we use a snapshot at an MHD statistical steady state from a
three dimensional simulation of supernova-driven turbulence as reported in
\cite{GMKS21}.  In a periodic Cartesian domain of 256 parsecs along each
dimension with mean gas number density of 1 cm$^{-3}$,  a weak random magnetic
field is amplified through dynamo to a mean energy density of about 5\%
equipartition with the mean kinetic energy density.  The model has a grid
resolution size of 0.5 parsecs along each edge.

We do not resolve self-gravity. The maximal cold gas number density minimally
of 600 K is a few tens cm$^{-3}$,  so that the corresponding Jeans length
$\lambda_{\rm J}$ of 43 parsecs exceeds the size of such structures in the
model.  Taking the mean density of 1 cm$^{-3}$ and a mean sound speed of
$12\kms$ in the warm gas,  $\lambda_{\rm J}>>256\pc$.  Effects of self-gravity
on dust cannot be entirely ruled out \cite{Mattsson22},  but we assess that
these effects are small in the present simulations,  in particular in the cases
that include the Lorentz force.  The maximal gas densities we can resolve are
constrained by the limited practicable resolution required to span a domain
size of $256\pc$ and adequately capture multiple SN remnants. The resulting
turbulence must be evolved tens of Myr to saturate a small-scale dynamo.
Increased resolution,  with increased maximal densities and reduced cooling
times,  increases by an order of magnitude both the size of each numerical
integration and the total integration time. Resolving molecular clouds,  is
therefore beyond the scope of this study. Similarly,  cooling by gas-dust
interaction is not taken into account because it would require that dust
physics is treated within the MHD simulation,  at a huge computational expense,
while the impact on the thermal sputtering rate is modest.

The turbulence is driven by SNe distributed uniform randomly in space at a
Poisson rate of approximately $1\Myr^{-1}$ within the domain. Equivalent
to a rate of around 20\% that of the Solar neighbourhood,  this lower rate
maintains a multiphase ISM with appropriate fractional volumes of cold,  warm
and hot gas.  In the periodic box,  with no escape for hot gas a higher SN rate
would quickly induce thermal runaway \cite{LOCBN15},  leaving the computational
domain saturated by hot gas and a tiny fraction of very dense cold and warm
gas.  Disk stratification with a halo into which hot gas can escape,  cool and
recirculate is omitted.

At the time at which we apply the dust-processing model to this simulation we
cease the continuous random SN explosions,  in order to isolate the effects
during the lifespan of a single SN remnant.  Using the same ambient state we
consider the case of two SN explosions,  one in a diffuse region of the ISM,
with $n$ approximately
0.03 cm$^{-3}$,  and another in a more dense region, with $n$ approximately 0.7 cm$^{-3}$.  We also run a control model,  in which no new SN is
added to the ambient state to isolate the dust processing induced by the
turbulent background dynamics from that of the SN blasts.

To model the explosion we inject $10^{51}\erg$ of thermal energy with spherical
Gaussian profile of radial scale 8 pc. At this resolution and ambient gas
density,  it is not necessary to include momentum injection \citep{KO15,
SBHO15} to obtain sufficient kinetic energy. In the highly dynamic and
inhomogeneous ISM the ideal analytic solutions of Sedov-Taylor do not apply,
but can be used \citep{GMKSH20} to verify the model rapidly evolves to match
the adiabatic solution within a few thousand years,  and well before
subsequently reproducing the snowplough evolution.  We omit SN mass ejecta,  as
the SN shockwave travels rapidly beyond the extent of the ejecta. The
snowplough phase and the interaction with the surrounding ISM has essentially
no connection with the properties of the ejecta,  which we can thus safely
ignore. The ambient density and flow are not altered to inject the SN,  so as
the flow evolves from the thermal pressure,  it immediately interacts with the
turbulent interior gas and magnetic field. Inevitably,  in the first few
thousand years dust destruction will be understated,  but this would be true
for all models,  including with the uniform ambient medium. Thus,  comparison
between our models provides a reliable indication of the relative effects
explored.

We further omit an evacuated bubble around the progenitor star created by
radiation pressure or stellar winds prior to the SN explosion. Cleared of most
of the gas and dust,  these bubbles and the surrounding wind-driven shells
extend in a homogeneous medium with $1$ particle per cm${}^{-3}$ up to 25 pc
after 1 Myr exposure time \cite{Martinez2019}. The blast wave needs only a few
kyr to reach these distances. The influence of these bubbles would affect only
a small fraction of our model domain and only the first <1 per cent of the
blast wave evolution time. Though the blast wave velocity and strength can be
disturbed  beyond that,  we expect a low impact of an evacuated bubble on the
total mass of destroyed dust.

For the generation of the multiphase MHD turbulence used for our ambient ISM
and its further evolution with or without our isolated SN explosion we solve
the set of nonideal compressible MHD equations,  using the sixth order Pencil
Code \cite{Pencil-JOSS} PDE solver.  As presented in \cite{KMG21} we include
radiative cooling and UV heating processes,  which apply rate of cooling
depending only on  a piece-wise varying exponent of temperature and normalised
by gas density. Cooling approximates the cumulative processes applying at
metallicity abundances expected in the Solar neighbourhood \cite{Wolfire:1995}.
This neglects the impact the dust could have on the thermodynamics of the gas,
if the dust together with a more sophisticated cooling dependent on the dust
abundances were included in the MHD model. It is reasonable to hypothesize,
from our results regarding dust-gas decoupling,  that such an effect would even
further reduce dust destruction,  as cooling would be even more effective where
dust density is highest.


In this study, following \cite{GMKS22}, we solve the set of nonideal MHD equations
  \begin{eqnarray}
  \label{eq:mass}
    \frac{\text{D}\rho}{\text{D}t} &=&
    -\rho \nabla \cdot \vect{u}
    +\nabla \cdot\zeta_D\nabla\rho,\\[0.5cm]
  \label{eq:mom}
    \rho\frac{\text{D}\vect{u}}{\text{D}t} &=&
    -\rho\cs^2\nabla\left({s}/{\cplocal}+\ln\rho\right)
    +\mu_0^{-1}\nabla\times\vect{B}\times\vect{B}
    \nonumber\\
    &+&\nabla\cdot \left(2\rho\nu{\mathbfss W}\right)
    +\rho\nabla\left(\zeta_{\nu}\nabla \cdot \vect{u} \right)
    \nonumber\\
    &+&\nabla\cdot \left(2\rho\nu_6{\mathbfss W}^{(5)}\right)
  -\vect u{\nabla}\cdot\left(\zeta_D{\nabla}\rho\right),\\[0.5cm]
  \label{eq:ind}
    \frac{\partial \vect{A}}{\partial t} &=&
    \vect{u}\times\vect{B}
    +\eta\nabla^2\vect{A}
    +\eta_6\nabla^6\vect{A},\\[0.5cm]
  \label{eq:ent}
    \rho T\frac{\text{D} s}{\text{D}t} &=&
    \EST\dot\sigma h^{-1} +\rho\Gamma-\rho^2\Lambda +\eta\mu_0^{-1} \vert\nabla \times \vect{B} \vert^2
    \nonumber\\
    &+&2 \rho \nu\vert{\mathbfss W}\vert^{2}
    +\rho~\zeta_{\nu}\left(\nabla \cdot \vect{u} \right)^2
    \nonumber\\
    &+&\nabla\cdot\left(\zeta_\chi\rho T\nabla s\right)
    +\rho T\chi_6\nabla^6 s
    \nonumber\\
    &-& \cv~T \left(
    \zeta_D\nabla^2\rho + \nabla\zeta_D\cdot\nabla\rho\right),
  \end{eqnarray}
and includes the ideal gas equation of statei, for which the adiabatic index is
$5/3$.  Treating the ISM as a monatomic, fully ionized plasma we apply a mean
molecular weight of 0.531.  Common variables and symbols take their usual
meanings.  $s$ is specific entropy and $\mathbfss W$ is the traceless rate of
train tensor, with $\mathbfss W^{(5)}$ its fifth order application to
hyperdiffusion.  Viscosity $\nu=5\cdot10^{-4}\kpc\kms$ and magnetic diffusivity
$\eta=10^{-4}\kpc\kms$. Shocks are resolved with artificial viscosities
$\zeta_D$, $\zeta_{\nu}$ and $\zeta_{\chi}\propto\nabla\cdot\vect{u}$, only
where flows are convergent. Sixth order hyperdiffusion applies coefficients
$\nu_6=\eta_6=\chi_6=6.25\cdot10^{-16}\kpc^5\kms$.
$\nabla^6=\partial^3_i\partial^3_i$.

$\EST=10^{51}\erg$ of SN energy are injected in equation\,\eqref{eq:ent} at a Poisson rate of about 60
$\kpc^{-3}\Myr^{-1}$.  Background ultarviolet heating is applied in equation\,\eqref{eq:ent} as
\begin{equation}\label{eq:Gamma}
\Gamma = \frac{\Gamma_0}{2}\left(1+\tanh\left[\frac{
     2\cdot 10^4\,{\rm K} - T}{2000\,{\rm K}}\right]\right),
\end{equation}
with $\Gamma_0 = 0.0147$\,erg\,g$^{-1}$\,s$^{-1}$.  The radiative losses
$\Lambda$ in equation\,\eqref{eq:ent} are modelled via a piece-wise power law dependence on temperature of
the form
\begin{equation}\label{eq:Lambda}
\Lambda(T) = \Lambda_k T^{\beta_k}\text{ for } T\in[T_k,T_{k+1}),
\end{equation}
with the parameters as listed in Table~\ref{tab:cooling}.

  \begin{table}
  \centering
    \caption{{\bf{The cooling function parameters equation~\eqref{eq:Lambda}}. The cooling coefficient $\Lambda_k$ applies to $T^{\beta_k}$ for temperature $T_k\leq T<T_k$.}
 \label{tab:cooling}}
\begin{tabular}{ccc}
\hline
 $T_k$             &$\Lambda_k$          &$\beta_k$           \\
$[{\rm K}]$        &$[\erg\,{\rm g}^{-2}\,{\rm s}^{-1}\,{\rm cm}^3\,{\rm K}^{-\beta_k}]$\\
\hline
 0                 & 0       & \phantom{$-$}$\cdots$ \\
 90                & 3.70e16 &   \phantom{$-$}2.12 \\
 141               & 9.46e18 &   \phantom{$-$}1.00 \\
 313               & 1.18e20 &   \phantom{$-$}0.56 \\
 6102              & 1.10e10 &   \phantom{$-$}3.21 \\
 1e5          & 1.24e27 &              $-0.20$\\
 2.88e5& 2.39e42 &              $-3.00$\\
 4.73e5& 4.00e26 &              $-0.22$\\
 2.11e6& 1.53e44 &              $-3.00$\\
 3.98e6& 1.61e22 &   \phantom{$-$}0.33 \\
 2.00e7& 9.23e20 &   \phantom{$-$}0.50 \\
 $\infty$ & $\cdots$ & \phantom{$-$}$\cdots$ \\
\hline
\end{tabular}
\end{table}


For the initial state of the dynamo simulation uniform ISM has gas number
density $n=1\cmcube$ and temperature $T=10^4$\,K. The seed field comprises
white noise with mean strength 1\,nanoGauss.  A snapshot from the saturated
state of the dynamo is used as the initial condition for all of the six cases
explored in this study.

We use the same MHD velocity field and gas structure to model the dust
processing with and without the Lorentz force arising from its magnetic field
for each case with an SN explosion and without.

\subsection*{Dust processing methods}\label{sec:dust_methods}

The evolution of the dust grains is driven by the gas conditions and by the
magnetic field in the turbulent,  inhomogeneous and shocked ISM. Pencil
provides snaphshots at intervals of 250\, yr from the 3D gas density,  gas
velocity,  gas temperature and magnetic field. 
We use our post-processing code \textsc{Paperboats}
\citep{Kirchschlager2019} to study the transport,  destruction and gas
accretion of the dust grains for these snaphshots. To follow the dust
evolution,  the ‘dusty-grid approach’ is used where the dust location is
discretized to spatial cells and the dust in each cell is apportioned in
different grain size bins. The dust grains can move both spatially as well as
between the grain size bins as a result of dust destruction or growth during a
time-step.

Due to the excessive computational effort required for highly resolved 3D
post-processing simulations,  we confine the study of dust-processing to a
thin slice through the center of the explosion,  spanning a cuboid volume of
$256\, \text{pc} \times 256\, \text{pc}\times 0.5\, \text{pc}$.  The
destroyed dust masses are computed for this thin slice and multiplied by 512
in order to estimate the destroyed dust masses of the entire 3D domain,
under an assumption that the dust processing is approximately isotropic.  For
the simulations including the blast wave the destroyed dust masses are scaled
by a correction factor of $0.365$,  which takes into account that the
explosion takes place in a central spherical region and that the blast wave
does not reach the outer edge of the domain within 1\, Myr. The value of
$0.365$ was derived by comparing the destroyed dust masses in the thin slice
of the homogeneous ambient ISM (as in \cite{KMG21}) to the inhomogeneous ISM,  and
expecting that the same ratio exists for the 3D domain between the
homogeneous and the inhomogeneous ISM.  We recognise that the remnant is far
from spherical and the turbulence far from isotropic in this system. The
slice is an arbitrary selection and it is reasonable to assume that it is
typical of the system.  When resources permit a stochastic study across
multiple slices and realisations of the ISM would better constrain these
assumptions.

In this study the dust transport, dust destruction and dust growth are determined. The acceleration of dust grains of mass $m$ occurs by gas-grain collisions,  by Coulomb interactions of charged grains in the ionised gas as well as by Lorentz forces on charged grains in magnetic fields \cite{Kirchschlager2023},
 \begin{align}
\mathbf{a}_\text{acc} = \frac{\mathbf{F}_\text{drag}}{m} + \frac{\mathbf{F}_\text{Lorentz}}{m}.
\end{align}
The net drag force caused by collisional drag and by plasma drag  (\cite{Baines65, Draine79}) is given as
\begin{align}
F_\text{drag} = 2 \sqrt{\pi} k_\text{B} T_\text{gas} a^{ 2} \sum_j n_{\text{gas},j} \left(\mathcal{F}_\text{col,j}+ \mathcal{F}_\text{pla,j}\right), \label{drag_force}
\end{align}
where $k_\text{B}$ is the Boltzmann constant, $T_\text{gas}$ is the gas temperature, $a$ is the grain size,  and $\mathcal{F}_\text{col,j}$  and  $\mathcal{F}_\text{pla,j}$ are the `Collisional term' and the `Plasma term', respectively. The sum runs over all plasma species $j$ within the gas (atoms, molecules, ions, and electrons). The Lorentz force is given by
\begin{align}
 \mathbf{F}_\text{Lorentz} =Q_\text{grain}\, \mathbf{v}_\text{rel}\times \mathbf{B}, \label{Lorentz}
\end{align}
where $Q_\text{grain}$ is the dust grain charge, $\mathbf{B}$ is the magnetic field, and $\mathbf{v}_\text{rel}$ is the relative velocity between dust grain and magnetic field. Dust grain charges are determined due to the ionisation of impinging plasma particles (ions and electrons), associated secondary electrons, transmitted plasma particles, and field emission \cite{Fry2020}.

The dust material can be destroyed by either sputtering or grain-grain collisions.  Sputtering \cite{Barlow78,Draine79, Dwek92} is the ejection of grain atoms due to the bombardment of gas particles (atoms, ions, or molecules).
We distinguish between kinetic sputtering (velocity of gas particles due to the grain moving relatively to the gas) and thermal sputtering (thermal motion of the gas particles).  The rate of decrease of grain radius $a$ per unit time, $\text{d}a/\text{d}t$, due to kinematic sputtering and thermal sputtering can be expressed as
\begin{align}
 \frac{\text{d}a}{\text{d}t} &= \frac{\left\langle M_\text{atom}\right\rangle}{2\,\rho_\text{bulk}}v_\text{rel} \sum_k n_{\text{gas},k} Y_k \left(E\right) \label{isput}\\
\text{and}\hspace*{0.5cm}&\nonumber\\
  \frac{\text{d}a}{\text{d}t} &= \frac{\left\langle M_\text{atom}\right\rangle}{2\,\rho_\text{bulk}}\sum_k n_{\text{gas},k} \left\langle Y_k v \right\rangle, \label{tsput}
\end{align}
respectively, where $\left\langle M_\text{atom}\right\rangle$ is the average atomic mass of the grain atoms, $\rho_\text{bulk}$ is the material density, $v_\text{rel}$ is the relative velocity between dust grains and the surrounding gas, $n_{\text{gas},k}$ is the number density of gas species $k$, $Y_k(E)$ is the sputtering yield of gas species $k$ as a function of the kinetic energy $E$, $\left\langle Y_k v \right\rangle$ is the sputtering yield averaged over the Maxwellian velocity distribution, and $v$ is the thermal velocity of a gas particle. For temperatures higher than $10^4\,$K the relative velocity between a grain and the surrounding gas is not unimodal but is a combination of the motion of the grain relative to the surrounding gas and the thermal motion of the gas particles. These two motions are combined using a skewed Maxwellian distribution instead of a regular Maxwellian distribution \cite{Shull78, Bocchio2014}. The size-dependent sputtering effect is also taken into account \cite{SerraDiazCano2008}.

The second important kind of dust destruction processes are grain-grain collisions. Collisions between dust grains of different sizes occur due to the relative velocities between them. The collision probability \cite{Kirchschlager2019} of a single grain of size $a_i$ to collide with any grain of size $a_j$ during the time interval $\Delta t$ is
\begin{align}
P_{ij}= 1-\exp{\left[-n_j\,\sigma_\text{col}\,v_\text{col}\,\Delta t\right]},
\end{align}
where $n_j$ is the number density of grains of size $a_j$, $\sigma_\text{col}$ is the collision cross section, and $v_\text{col}$ is the collision velocity. Both $\sigma_\text{col}$ and  $v_\text{col}$ take repulsion or attraction due to Coulomb interaction between the charged dust grains into account.  The outcome of a grain-grain collision \cite{Borkowski95,
Jones96} depends on the collision energy. For the largest collision
velocities (>19 km/s),  the dust grains can be fully vaporized which means
that the whole material goes into the gas phase. Please note that a partial
vaporization approach as in \cite{Kirchschlager2023} is not included but has
shown to be mainly important for sizes larger than the MRN
\cite{Mathis77} grains. On the other hand, intermediate collision velocities result in
(partial) shattering of the dust grains \cite{Hirashita09}. The shattered material is then
redistributed in a size distribution of fragments. At collision velocities
below 2.7\, km/s,  the grains are not shattered but bounce or even stick
together,  though the latter process has a low occurrence in the simulations
as the gas and dust velocities are too high. Further dust growth processes
like gas accretion or ion trapping \cite{Kirchschlager2020} of destroyed dust
material also play only a minor role. We neglect gas accretion and ion trapping of the regular gas due to the nature of post-processing \cite{Kirchschlager2020}.

Dust processing with and without the Lorentz force acting on
the dust is considered to assess its effect on the dust evolution. In either case the gas
velocities and structure are obtained from the same MHD simulation.  The
initial spatial distribution of the dust follows the gas,  assuming a constant
gas-to-dust-mass ratio of 100. The grains are made of silicate and follow
initially a size distribution of MRN type. The grains are
binned in 20 size bins with an additional collector bin at the lower and upper
end of the size distribution,  respectively.

It is easier to isolate the location of SN explosions modelled in a uniform
ambient ISM in order to calculate the dust processing associated with the
remnant. In such an inhomogenous ambient ISM as applies here,  this is
difficult to do.  We,  therefore,  include a model without an SN explosion,  so
we can subtract the dust processing which occurs absent the SN,
background processing.  While the MHD model has a mean gas number
density of 1 cm$^{-3}$,  due to its turbulent evolution the initial mean
density of the two-dimensional slice on which we model the dust processing is
0.59\, cm$^{-3}$.  For the duration of the dust processing models,  the mean
gas density has negligible monotonic decay (\mbox{Supplementary~Fig.~7}),  but
locally can fluctuate modestly due to the advection of gas in and out of the
plane within the three-dimensional turbulence.

Our results of purely background processing show that a statistical steady
state filamentary dust distribution evolves from the initial state within about
300 kyr. Dust processing in the early aftermath of an explosion in future
simulations could,  therefore,  be improved by evolving the dust processing 300
kyr before exploding the SN.

\subsection*{Data availability}

The Pencil Code simulation run files and the initial snapshot used at the
beginning of the dust processing experiment (about 20 GB) have been deposited
in the Finnish Fairdata storage service under the accession code
\url{https://doi.org/10.23729/ac4542ad-ab85-4ccc-8d0c-60bfc5472ef2}
from which the MHD solutions can be reproduced without having to replicate the
dynamo simulations. The start and run files are also included for replicating
the dynamo simulations.

The time series of 2D slices alone of gas density, temperature, velocity
and magnetic field, and the subsequent slices of dust densities used in the
analysis exceed 1.4TB. Public hosting is impractical for the authors due to
this large size. However, the datasets generated and analysed during
the current study are available from the corresponding author on request.

\subsection*{Code availability}

We use the Pencil Code \cite{Pencil-JOSS} to perform all simulations,  which
is freely available under \url{https://github.com/pencil-code/}
\url{https://doi.org/10.21105/joss.02807}.  The Paperboats code
\cite{Kirchschlager2019, Kirchschlager2020, KMG21, Kirchschlager2023} is
available at \url{https://doi.org/10.5281/zenodo.10036806}.

\subsection*{Author contributions statement}
F.K., L.M. and F.G. have made essential contributions to the conceptualisation,  data
analysis and writing of this paper.

\subsection*{Competing interests statement}
The authors declare no competing interests.



\subsection*{Acknowledgments}

We acknowledge funding from the European Research Council (ERC) under the
European Union’s Horizon 2020 research and innovation programs:
SNDUST ERC-2015-AdG-694520 (F.K.),
DustOrigin ERC-2019-StG-851622 (F.K)
and UniSDyn  grant no. 818665 (F.G.);
the Swedish Research Council (Vetenskapsr\aa
det), grants no. 2015-04505 (L.M.) and 2022-03767 (L.M.);
the Academy of Finland ReSoLVE Centre of Excellence grant 307411 (F.G.); and
the Ministry of Education and Culture Global Programme USA Pilot 9758121 (F.G.).
We appreciate the generous computational resources from CSC -- IT Center for
Science, Finland, under Grand Challenge GDYNS Project 2001062.

\newpage

\end{document}